\documentclass[journal]{IEEEtran}
\usepackage{amsmath,amsfonts}
\usepackage{amssymb}
\usepackage[ruled]{algorithm2e}
\usepackage[noend]{algpseudocode} 
\usepackage{array}
\usepackage{algorithmicx}    
\usepackage{threeparttable}
\usepackage{multirow}
\usepackage{diagbox} 
\usepackage{subfigure}
\usepackage{xcolor}
\usepackage{multicol}   
\usepackage{multirow}    
\usepackage{float}  
\usepackage{makecell}  
\usepackage{booktabs}
\usepackage{utfsym}
\usepackage{textcomp}
\usepackage{stfloats}
\usepackage{url}
\usepackage{verbatim}
\usepackage{graphicx}
\usepackage{cite}

\begin{document}

\title{\makebox[\linewidth]{\parbox{\dimexpr\textwidth+2cm\relax}{\centering Deep Learning-based CSI Feedback in Wi-Fi Systems}}}

\author{{\IEEEauthorblockN{
Fan~Qi\IEEEauthorrefmark{1},
Jiajia~Guo\IEEEauthorrefmark{1},
Yiming~Cui\IEEEauthorrefmark{1},
Xiangyi~Li\IEEEauthorrefmark{1},
Chao-Kai~Wen\IEEEauthorrefmark{2},
and
Shi~Jin\IEEEauthorrefmark{1} }

		\IEEEauthorblockA{\IEEEauthorrefmark{1}National Mobile Communications Research Laboratory, Southeast University, Nanjing 210096, P. R. China,\\Email:
		\{qifan, jiajiaguo, cuiyiming, Xiangyi\_li, jinshi\}@seu.edu.cn}
		
		\IEEEauthorblockA{\IEEEauthorrefmark{2}Institute of Communications Engineering, National Sun Yat-sen University, Kaohsiung 80424, Taiwan, \\Email:  chaokai.wen@mail.nsysu.edu.tw}
     
	}
 
\thanks{ 	
This work has been submitted to the IEEE for possible publication. Copyright may be transferred without notice, after which this version may no longer be accessible.
}
}

\maketitle

\begin{abstract}

In Wi-Fi systems, channel state information (CSI) plays a crucial role in enabling access points to execute beamforming operations. However, the feedback overhead associated with CSI significantly hampers the throughput improvements. Recent advancements in deep learning (DL) have transformed the approach to CSI feedback in cellular systems. Drawing inspiration from the successes witnessed in the realm of mobile communications, this paper introduces a DL-based CSI feedback framework, named EFNet, tailored for Wi-Fi systems.
The proposed framework leverages an autoencoder to achieve precise feedback with minimal overhead. The process involves the station utilizing the encoder to compress and quantize a series of matrices into codeword bit streams, which are then fed back to the access point. Subsequently, the decoder installed at the AP reconstructs beamforming matrices from these bit streams.
We implement the EFNet system using standard Wi-Fi equipment operating in the 2.4 GHz band. Experimental findings in an office environment reveal a remarkable 80.77\% reduction in feedback overhead compared to the 802.11ac standard, alongside a significant boost in net throughput of up to 30.72\%.

\end{abstract}


\vspace{1mm}
\section{Introduction}

As wireless communication technology advances swiftly and the need for bandwidth grows, Wi-Fi deployment in various settings like enterprises, campuses, and residences is becoming increasingly prevalent. The proliferation of Wi-Fi hotspots and the continuous evolution of Wi-Fi systems have escalated the expectations for key performance indicators such as spectrum efficiency, end-to-end latency, and data rates. The inception of multiple-input multiple-output (MIMO) technology in the 802.11n standard marked a significant leap in data transfer rates over its predecessors\cite{IEEE802.11n}.

To enable beamforming, the access point (AP) relies on channel state information (CSI) to craft beamforming filters for subsequent data transmission. The AP can obtain CSI through two methods: explicit and implicit feedback. Explicit feedback involves the station (STA) directly transmitting estimated beamforming matrices to the AP in compressed or uncompressed formats. In contrast, implicit feedback leverages channel reciprocity to deduce CSI from received signals and various parameters, demanding sophisticated channel calibration algorithms. With the exclusion of implicit feedback in the IEEE 802.11ac standard\cite{IEEE802.11ac}, the adoption of explicit feedback becomes imperative. Nonetheless, Wi-Fi's use of orthogonal frequency division multiplexing (OFDM) modulation generates multiple subcarriers, each corresponding to a channel matrix, resulting in substantial feedback overhead.

In recent years, there has been rapid development in the field of artificial intelligence (AI). As a significant technology in AI, deep learning (DL) has been widely applied in wireless communication. CsiNet was the first DL-based CSI feedback architecture proposed in mobile communication\cite{wen2018deep}. It can learn an effective way to compress and reconstruct CSI with higher reconstruction quality compared with conventional methods. Since then, research on the combination of DL and CSI feedback has emerged subsequently in mobile communication\cite{yang2019deep,guo2022overview,tang2022dilated,bi2022novel}.

However, the models mentioned cannot be directly applied from the cellular network scenario to the Wi-Fi system due to noticeable disparities between the CSI in Wi-Fi and cellular environments. These differences manifest in three key aspects:
\begin{enumerate}
    \item \textbf{Dimensions:} CSI dimensions in Wi-Fi setups are typically smaller than those in cellular networks and may lack clear sparsity and compressibility. Consequently, the effectiveness of DL-based data-driven models is not guaranteed.

    \item \textbf{System Reliability:} Most of the aforementioned works rely on simulated CSI data, which may not accurately reflect real-system Wi-Fi network performance, thus questioning the validity of the model.

    \item \textbf{Evaluation Metrics:} Different from the Wi-Fi system, evaluations in the cellular domain often overlook throughput, as the performance gain of sufficiently high reconstruction accuracy (NMSE below -10 dB) may not markedly enhance system throughput.
\end{enumerate}

To enhance the application of AI in Wi-Fi networks, the IEEE initiated the IEEE 802.11 AI/ML Topical Interest Group (TIG) in May 2022 \cite{giordano2023will}. The IEEE 802.11 AI/ML TIG Technical Report Draft \cite{wang2024} outlines five use cases, including CSI feedback compression, which employs AI/ML to boost Wi-Fi performance. Nonetheless, AI-based CSI feedback implementations in Wi-Fi systems remain relatively limited \cite{sangdeh2020lb,deshmukh2022intelligent,sangdeh2021deepmux,10272483}. Notably, LB-SciFi \cite{sangdeh2020lb} employs two autoencoders to compress two sets of angles from the beamforming matrix via Givens rotations, adhering to IEEE 802.11ac/ax standards \cite{IEEE802.11ac,IEEE802.11ax}. This method, however, requires complex data preprocessing and heavily depends on manual bit allocation across the two sets of angles.

In contrast to the CSI feedback mechanisms in IEEE 802.11ac/ax, this paper introduces a new DL-based explicit CSI feedback framework for Wi-Fi systems, named EFNet. EFNet employs neural networks to \emph{directly} learn and extract CSI representations within the Wi-Fi system. This approach addresses the excessive dependency on angular distributions observed in LB-SciFi, eliminates the need for manual preprocessing and bit allocation, and facilitates a fully automated CSI compression and reconstruction process. EFNet uses an autoencoder to exploit the correlations in the frequency domain of beamforming matrices, compressing them into a concise set of codewords via the encoder, which are subsequently reconstructed by the decoder. Experimental evaluations demonstrate that EFNet significantly reduces feedback overhead and enhances system performance compared to the standard and the LB-SciFi model.

\vspace{1mm}
\section{System model}
\label{section:system model}
In this section, the MIMO system model in Wi-Fi systems is introduced, which is followed by the explicit CSI feedback in IEEE 802.11ac/ax\cite{IEEE802.11ac,IEEE802.11ax} as well as in LB-SciFi.

\subsection{MIMO Protocol in Wi-Fi}

\begin{figure}[t]
    \centering
    \setlength{\abovecaptionskip}{-2mm}
    \includegraphics[width=0.95\linewidth]{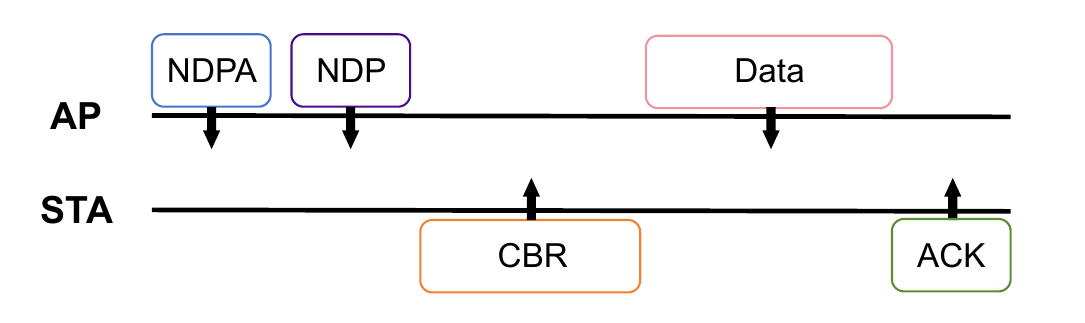}

    \caption{Illustration of MIMO protocol in Wi-Fi systems.}
    \label{standard}
\vspace{-0.1cm}
\end{figure}

This paper considers a downlink MIMO transmission system, where the AP is equipped with $N_{\mathrm{t}}$ transmit antennas and the STA is equipped with $N_{\mathrm{r}}$ receive antennas. In Wi-Fi systems, as illustrated in Fig.~\ref{standard}, in order to obtain CSI, the AP initiates a sounding sequence by first transmitting the null data packet announcement (NDPA) frame. The NDPA frame carries all necessary probing details to notify the STA of the MIMO transmission, such as beamforming training, address, and bandwidth information. Subsequently, the AP sends an null data packet (NDP) to prompt the STA to estimate the downlink CSI.
Then, the STA feeds the compressed CSI back to the AP, which is included in the compressed beamforming report (CBR). Upon receiving the CSI, the AP initiates beamforming operations and proceeds with downlink data transmission. Upon decoding the data packet, the STA sends an ACK or NACK to the AP to indicate whether packet detection is successful or unsuccessful \cite{IEEE802.11ax}.

In practical systems, the STA needs to transmit CSI to the AP continuously for subsequent beamforming. However, with the escalation in the number of antennas, the feedback of full channel matrices may engender considerable overhead. This, in turn, can lead to an augmentation in the transmission time of the CBR, thereby diminishing system throughput.
To mitigate the transmission time of CSI, the IEEE 802.11 standard employs compressed CSI, which is based on angles instead of the original beamforming matrices for feedback. This approach will be introduced in detail in the next subsection.

\subsection{Explicit Feedback in IEEE 802.11 Standard}

As mentioned above, the AP send NDP ($\mathbf{x}[k]$) to the STA for CSI. The received NDP by the STA over the $k$-th subcarrier can be represented as:
\begin{equation}
\mathbf{y}[k] = \mathbf{H}[k]\mathbf{x}[k] + \mathbf{n}[k], \quad k = 0, 1, \ldots, N_\mathrm{fft}-1,
\end{equation}
where $\mathbf{y}[k]\in\mathbb{C}^{N_\mathrm{r}\times 1}$, $\mathbf{H}[k]\in\mathbb{C}^{N_\mathrm{r}\times N_\mathrm{t}}$, $\mathbf{x}[k]\in\mathbb{C}^{N_\mathrm{t}\times 1}$, and $\mathbf{n}[k]\in\mathbb{C}^{N_\mathrm{r}\times 1}$ represent the received NDP, channel matrix, transmitted NDP, and white Gaussian noise vector respectively. $N_\mathrm{fft}$ represents the total number of subcarriers in an OFDM symbol.

The STA feedbacks two sets of angles which are decomposed instead of the raw CSI for each subcarrier. Firstly, perform singular value decomposition (SVD) to $\mathbf{H}[k]$:
\begin{equation}
\mathbf{H}[k]=\mathbf{U}[k]\boldsymbol{\Sigma}[k]\bar{\mathbf{V}}^\mathrm{H}[k],
\end{equation}
where $\mathbf{U}[k]\in\mathbb{C}^{N_\mathrm{r}\times N_\mathrm{r}}\text{ and }\bar{\mathbf{V}}[k]\in\mathbb{C}^{N_\mathrm{t}\times N_\mathrm{t}}$, which are the unitary matrices, and $\boldsymbol{\Sigma}[k]\in\mathbb{R}^{N_\mathrm{r}\times N_\mathrm{t}}$ is a diagonal matrix. The diagonal elements of $\boldsymbol{\Sigma}[k]$ are a series of decreasing singular values of $\mathbf{H}$, denoted as $\{\sigma_i,i=1,\ldots,N_{\min}\}$, where $N_{\mathrm{min}}=\min(N_\mathrm{r},N_\mathrm{t})$.

For purpose of matrix dimension reduction, the data of the first $N_\mathrm{s}$ columns of $\bar{\mathbf{V}}[k]$ are first extracted, which is represented as $\mathbf{V}[k]$, where $N_\mathrm{s}$ is the number of data stream. Then, Givens rotation is presented to $\mathbf{V}[k]$, which can be expressed by:
\begin{equation}
\mathbf{V}[k]=\left(\prod_{i=1}^{\min(N_\mathrm{s},N_\mathrm{t}-1)}\left(\mathbf{D}_i(\phi^{(k)})\prod_{l=i+1}^{N_\mathrm{t}-1}\mathbf{G}_{li}^\mathrm{T}(\psi_{li}^{(k)})\right)\right)\mathbf{I}_{N_\mathrm{t}\times N_\mathrm{s}}.
\end{equation}
In the formula,  $\mathbf{D}_i(\phi)$ is an $N_\mathrm{t}\times N_\mathrm{t}$ diagonal matrix:
\begin{equation} 
\left.\mathbf{D}_i(\phi)=\left[\begin{array}{ccccc}\mathbf{I}_{i-1}&0&0&\ldots&0\\0&e^{j\phi_{i,i}}&0&\ldots&0\\0&0&\ddots&0&0\\0&0&0&e^{j\phi_{N_\mathrm{r}-1,i}}&0\\0&0&0&0&1\end{array}\right.\right],
\end{equation}
and $\mathbf{G}_{li}(\psi)$  is an $ N_\mathrm{t}\times N_\mathrm{t}$ Givens rotation matrix:
\begin{equation}
\left.\mathbf{G}_{li}(\psi)=\left[\begin{array}{ccccc}\mathbf{I}_{i-1}&0&0&0&0\\0&\cos\psi&0&\sin\psi&0\\0&0&\mathbf{I}_{l-i-1}&0&0\\0&-\sin\psi&0&\cos\psi&0\\0&0&0&0&\mathbf{I}_{N_\textbf{r}-l}\end{array}\right.\right],
\end{equation}
where $\mathbf{I}_{m}$ represents an $m\times m$ identity matrix, and $\mathbf{I}_{N_\mathrm{t}\times N_\mathrm{s}}$ is an $N_\mathrm{t}\times N_\mathrm{s}$ identity matrix padded with 0s to fill the additional rows or columns when $N_\mathrm{t} \neq N_\mathrm{s}$. 
 
Due to the high correlation of CSI among neighboring subcarriers, an STA can group multiple neighboring subcarriers together for CSI feedback instead of providing CSI feedback for each subcarrier individually. According to the IEEE 802.11ac standard, the number of subcarriers in each group (denoted as $N_\mathrm{g}$) can be 1, 2, or 4.

Through these operations, two sets of angles, $\mathbf{\Psi}$ and $\mathbf{\Phi}$, are obtained which are used by the STA for feedback. The number of angles in each set increases as the value of ${N_\mathrm{t}}$ and ${N_\mathrm{r}}$ increases, leading to a corresponding increase in feedback overhead. To reduce the overhead, the standard proposes two quantization schemes for $\mathbf{\Psi}$ and $\mathbf{\Phi}$. Specifically, the number of quantization bits in Type $0$ are 5 and 7 for angles in $\mathbf{\Psi}$ and $\mathbf{\Phi}$ respectively, while in Type $1$ they are 7 and 9. Type $1$ improves the feedback performance at the expense of more overhead. Upon receiving the feedback from the STA, the AP reconstructs $\mathbf{V}$ through inverse operations. The reconstructed $\mathbf{V}$ matrices are used for subsequent beamforming.

\begin{figure}[t]
    \centering
    \setlength{\abovecaptionskip}{-2mm}
    \includegraphics[width=0.95\linewidth]{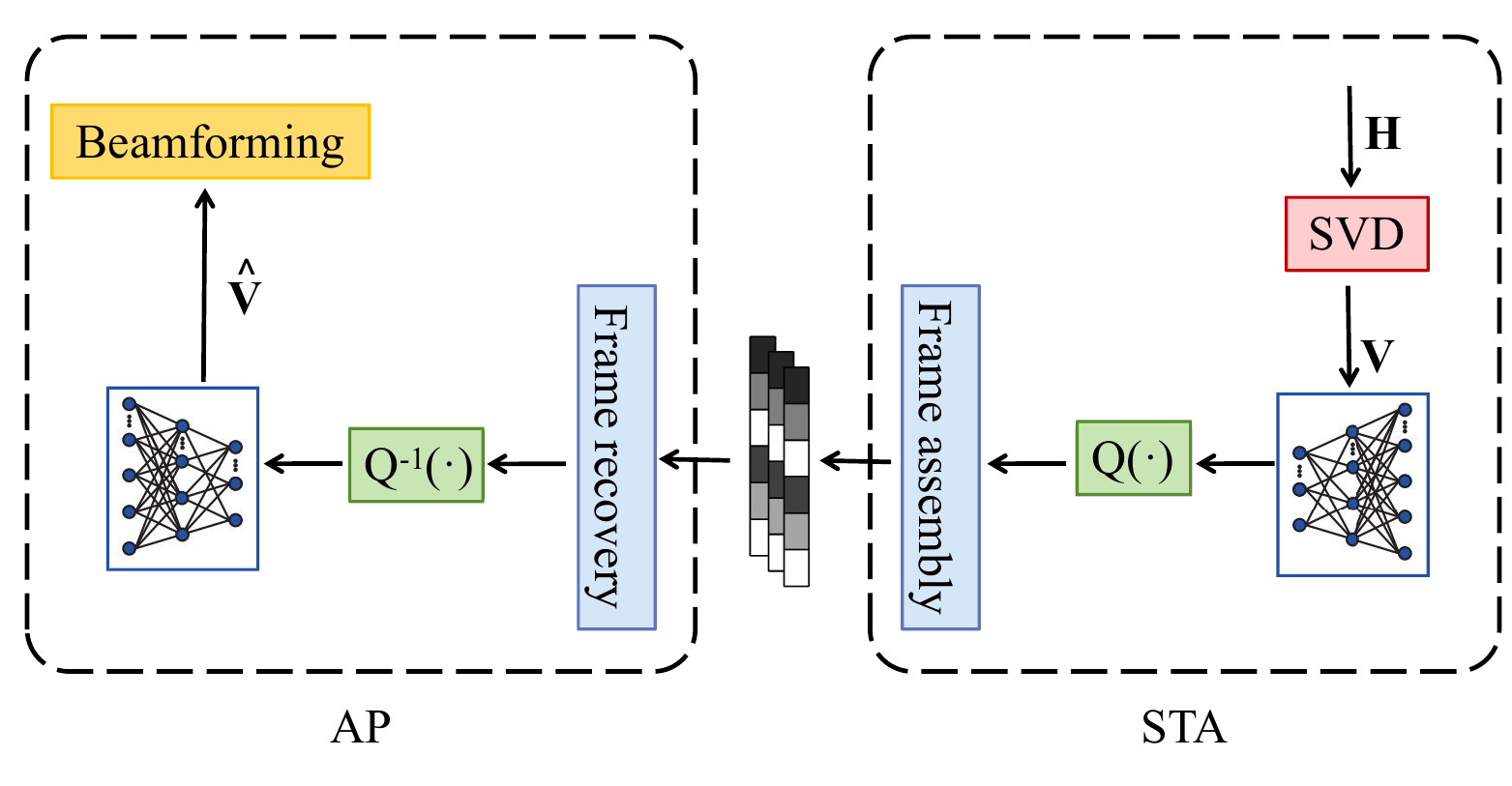}
    \caption{Illustration of EFNet framework in Wi-Fi systems. The proposed EFNet framework includes the encoder at the STA and the decoder at the AP, which compresses and recovers the beamforming matrices $\mathbf{V}$ respectively.}
    \label{feedback}
\vspace{-0.1cm}
\end{figure}

\subsection{Explicit Feedback in LB-SciFi} 

LB-SciFi \cite{sangdeh2020lb} is the first learning-based CSI feedback framework proposed for Wi-Fi systems. It features two DNN-AEs that compress CSI at each STA and reconstruct it at the AP. The framework is structured into two phases: online training and real-time exploitation. During online training, the AP uses side information ($\psi$ and $\phi$) derived from existing 802.11 protocols to train the DNN-AEs. After training, the AP broadcasts the encoders to all STAs, initiating the exploitation phase where these encoders are used for CSI feedback.

The distinctiveness of angle sets $\psi$ and $\phi$ in terms of their power spectral entropy (PSE) was investigated to determine if the same DNN-AE configuration could be applied to both. Results showed that $\psi$ and $\phi$ exhibit different PSE values, indicating varied compressibility. Consequently, this necessitated the use of two different DNN-AEs. Subsequent trials focused on optimizing the bit allocation for the code dimension of each DNN-AE, selecting dimensions based on a predetermined reference compression error.

Prior to training, datasets require preprocessing to ensure that the data has a normalized zero-mean PDF and uniform variance across subcarriers \cite{kim1999normalization}. This involves applying a linear transformation to $\psi$ and a piecewise linear transformation to $\phi$. Such preprocessing is critical to avoid biased training outcomes and to accelerate the convergence of the DNN-AEs. However, it is important to note that preprocessing parameters must be tailored to each specific dataset.

\vspace{1mm}
\section{EFNET: DL-BASED EXPLICIT FEEDBACK ARCHITECTURE}\label{section:EFNet}

In this section, we will first discuss the motivation behind designing EFNet, the DL-based explicit feedback architecture. Then, we will introduce the main framework of EFNet, followed by a description of the neural network design.

\begin{figure*}[t]
    \centering
    \setlength{\abovecaptionskip}{-2mm}
    \includegraphics[width=0.9\linewidth]{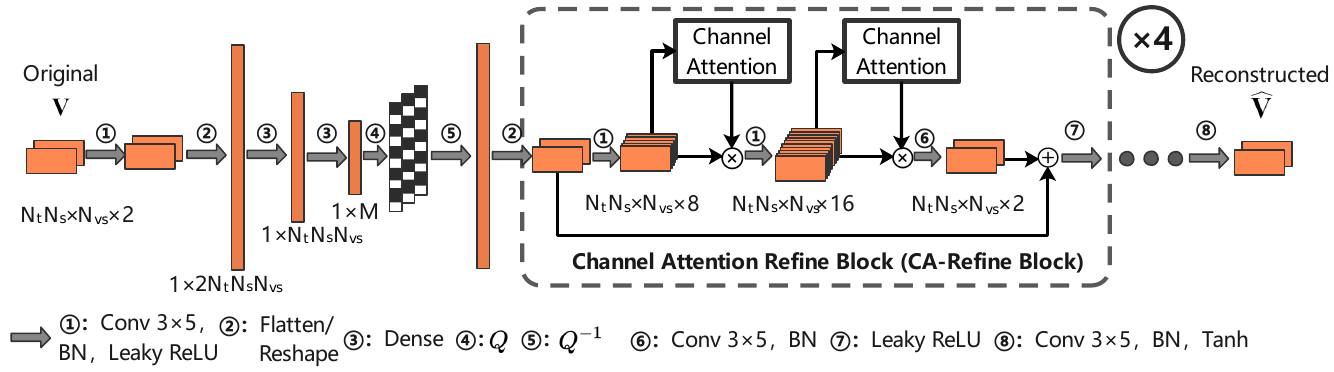}

    \caption{The detailed neural network design of the proposed EFNet inspired by CsiNet\cite{wen2018deep}.}
    \label{efnet}
\vspace{-0.4cm}
\end{figure*}

\begin{figure}[t]
    \centering
    \setlength{\abovecaptionskip}{-2mm}
    \includegraphics[width=0.9\linewidth]{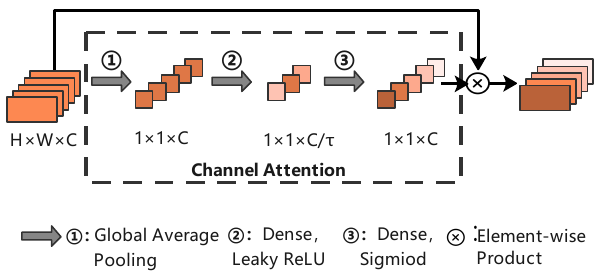}

    \caption{The architecture and mechanism of Channel attention module.}
    \label{Attention}
\vspace{-0.4cm}
\end{figure}

\subsection{Motivation}

LB-SciFi, as introduced earlier, presents inherent shortcomings. The necessity for specific precoding strategies, contingent on the CSI distribution, mandates manual bit allocation for dual sets of angles during the design of codeword quantization schemes. Data distribution intricacies not only influence the configurations of LB-SciFi and the structural parameters of neural networks but also introduce complexities in aligning with established standards. Furthermore, environmental shifts, such as alterations in channel characteristics, threaten CSI reconstruction performance when utilizing existing precoding techniques and network configurations. Consequently, the synchronous modification of precoding and bit allocation schemes becomes imperative, demanding extensive human intervention. This process also risks suboptimal bit allocation decisions, underscoring the need for further refinement in this aspect.

To address the challenges mentioned earlier, we propose a DL-based explicit CSI feedback framework. Our approach involves extracting multiple channel matrices from a single data packet and combining them to create an ``image'' post singular value decomposition (SVD). Subsequently, an autoencoder is trained using numerous ``image'' collected from consecutive data packets, aiming to capture the characteristics of CSI across adjacent subcarriers. This methodology harnesses the capabilities of DL by facilitating the automatic learning of the optimal bit allocation scheme during the training process. Unlike LB-SciFi's preprocessing and manual bit allocation requirements, our proposed approach eliminates the need for prior data analysis before inputting the data into the neural network. This streamlined process allows for efficient feature learning of CSI attributes, enhancing the overall effectiveness of the feedback framework.

\subsection{EFNet Framework}

The workflow of the proposed EFNet shown in Fig. \ref{feedback} is as follows.
Firstly, a series of matrices $\{\mathbf{H}[k]\}_{k=0}^{N_{\mathrm{fft}}-1}$ matrices estimated by the STA from one data packet for all subcarriers are put into a SVD operation to obtain $\{\Bar{\mathbf{V}}[k]\}_{k=0}^{N_{\mathrm{fft}}-1}$. For dimensional reduction, we extract the first $N_\mathrm{s}$ columns for each $\Bar{\mathbf{V}}[k]$ and extract $N_{\mathrm{vs}}$ valid subcarriers with equal spacing, denoted as $\{\mathbf{V}[k]\}_{k=0}^{N_{\mathrm{vs}}-1}$.
Subsequently, we vectorize $\mathbf{V}[k]$ into a shape of $N_\mathrm{t} N_\mathrm{s}\times 1$ and combine $\{\mathbf{V}[k]\}_{k=0}^{N_{\mathrm{vs}}-1}$ into a two-dimensional ``image" with a shape of $N_\mathrm{t} N_\mathrm{s}\times N_\mathrm{vs}$. Next, the complex-valued elements are transformed into real values (real and imaginary parts) and normalized into a range of $[-1,1]$. This $N_\mathrm{t} N_\mathrm{s}\times N_\mathrm{vs} \times 2$-dimensional ``image", denoted as $\mathbf{V}$ in Fig. \ref{feedback}, is sent into the encoder for compression.
Following this, the compressed codewords are quantized to bits considering the practical transmission of data is in the form of bits. The bit-stream codeword $\mathbf{s}_{\mathrm{bit}}$ can be given as:
\begin{equation}
    \mathbf{s}_{\mathrm{bit}}=Q(f_{\mathrm{Enc}}(\mathbf{V};\Pi_{\mathrm{Enc}})),
\end{equation} 
where $f_{\mathrm{Enc}}(\cdot)$, $\Pi_{\mathrm{Enc}}$ and $Q(\cdot)$ represent the encoder's function, the corresponding network parameters and the uniform quantization operation with $q$ quantization bits, respectively. Then the feedback bitstreams are assembled into frames for data transmission.

Upon receiving the data sent by the STA, the AP extracts the bitstreams and dequantizes them to codewords. The decoder is utilized to reconstruct $\mathbf{V}$, which will be used for subsequent beamforming. The reconstructed $\hat{\mathbf{V}}$ is described as:
\begin{equation}
    \hat{\mathbf{V}}=f_{\mathrm{Dec}}(Q^{-1}(\mathbf{s}_{\mathrm{bit}});\Pi_{\mathrm{Dec}}),
\end{equation}
where $f_{\mathrm{Dec}}(\cdot)$, $\Pi_{\mathrm{Dec}}$ signify the decoder and its corresponding parameters, respectively.

\subsection{Network Architecture}
The overview of the designed network of EFNet is shown in Fig. \ref{efnet}.
As the encoder's input, the beamforming matrix $\mathbf{V}$ is first put into a convolutional layer with a kernel of $3 \times 5$, then reshaped into a vector, followed by a fully connected layer to generate the real-valued $M$-length codeword. The codeword is then fed to the decoder, which uses a fully connected layer and a reshaping layer to reconstruct its original dimension. The coarse-grained CSI reconstructed feature maps are fed to four Channel Attention Refine Blocks (CA-Refine Blocks) to refine and improve the reconstruction. Each CA-Refine Block consists of two channel attention networks, the architecture of which is shown in Fig. \ref{Attention}.

The channel attention mechanism employs a learnable neural network to predict the importance of each channel, allowing for weighting these channels based on their respective importance to enhance the network's perception of critical features. Upon obtaining $H \times W \times C$ feature maps, we employ global average pooling to obtain a $1 \times 1 \times C$ vector. Subsequently, a fully connected layer is utilized to transform this vector into a $C/\tau$-dimensional vector, where $\tau$ represents a reduction factor, with a Leaky ReLU activation function. We set $\tau = 2$ in our experiments. Following this, another fully connected layer is employed to reconstruct this $C/\tau$-dimensional vector into a $C$-dimensional vector, with a sigmoid activation function to output the normalized weights. This $C$-dimensional vector is then utilized to perform element-wise multiplication with the convolutional feature maps, resulting in the final feature maps. The differential attention weights on various feature map channels are determined by distinct $\mathbf{V}$ matrices. This mechanism enables the network to extract more salient information, thereby facilitating improved reconstruction of $\mathbf{V}$ matrices.

Zero padding is used to maintain the same size matrix for each layer. The final convolutional layer with a $\tanh$ activation function outputs the reconstructed beamforming matrix. The ADAM\cite{diederik2014adam} algorithm is chosen for the optimization of the training parameters, and the mean square error (MSE) loss function is given as:
\begin{equation} 
     \mathcal{L}_{\mathrm{MSE}}(\Pi|\mathcal{D})=\frac{1}{|\mathcal{D}|}\sum_{\mathbf{V}\in\mathcal{D}} \|\hat{\mathbf{V}}-\mathbf{V} \|_{2}^{2},
\end{equation}
where the dataset $\mathcal{D}$ comprises of $\mathbf{V}$ matrices, and $|\mathcal{D}|$ denotes the cardinality of set $\mathcal{D}$, signifying the total number of samples contained within the dataset.

\vspace{1mm}
\section{Experiment and Results}\label{section:experiment}

\subsection{Experimental Setup and Preparation}

\begin{figure}[t]
    \centering
 
	\includegraphics[width=1\linewidth,trim=1.5cm 0cm 1cm 0cm,
	clip]{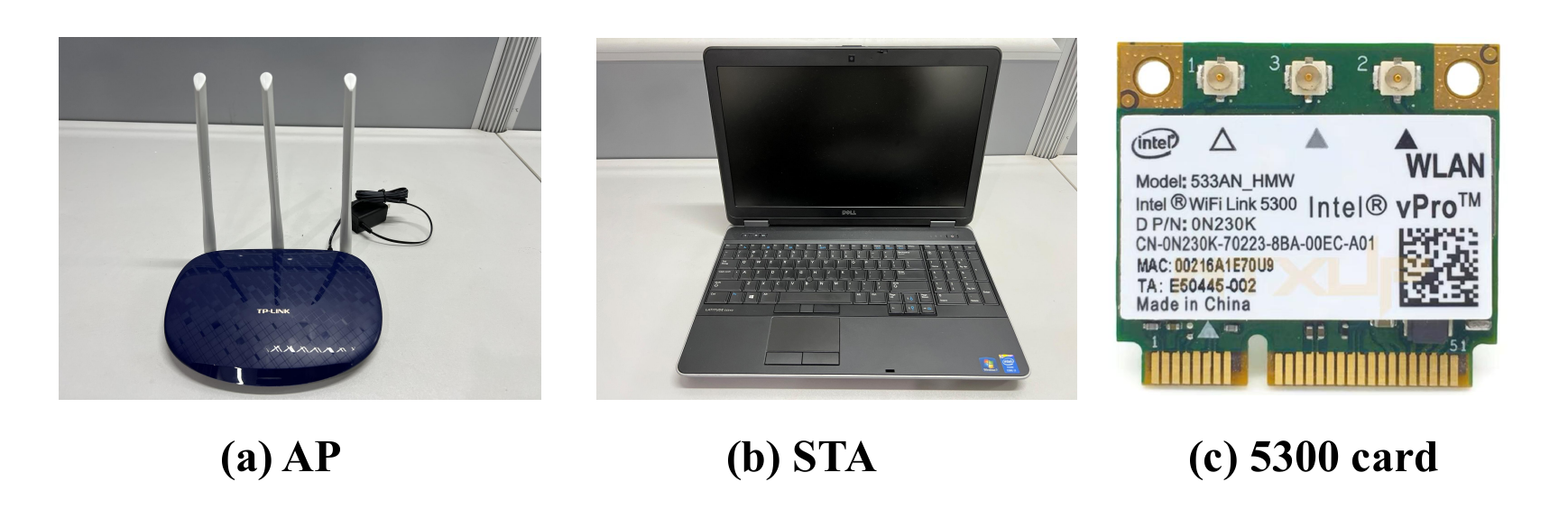}
	\setlength{\abovecaptionskip}{-0.2cm}

    \caption{Devices used for the measurement.}
    \label{devices}
\vspace{-0.1cm}
\end{figure}

We assess the efficacy of the proposed DL-based EFNet by conducting practical experiments, juxtaposing its performance against that of the 802.11 standards and LB-SciFi. The experimental setup includes a DELL E6540 laptop, a TP-LINK TL-WR886N wireless router, an Intel 5300AGN wireless network card, and the configuration of the CSI Tool platform\cite{halperin2011tool}. The wireless router, equipped with three antennas, functions as the AP, whereas the laptop serves as the STA. The laptop is powered by an Intel Core i7-2640 CPU, operating Ubuntu 16.04 LTS with Linux kernel version 4.2.1. Integrated into the laptop is an Intel 5300AGN wireless network card with three antennas. The physical setup of the devices is shown in Fig. \ref{devices}.

\begin{figure}[t]
    \centering

    \includegraphics[width=1\linewidth]{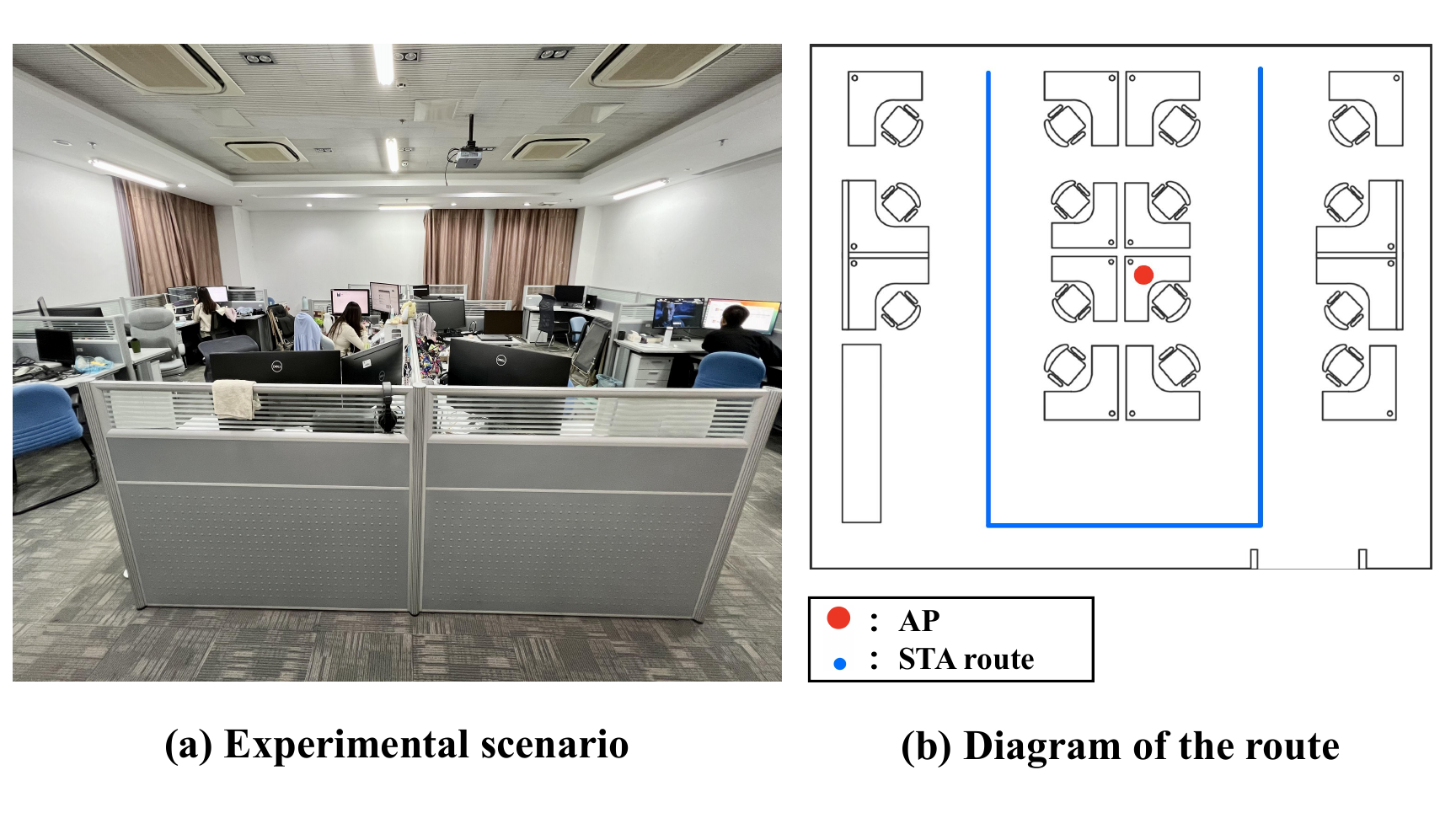}
    \setlength{\abovecaptionskip}{-0.2cm}
    \caption{Scenarios for the measurement.}
    \label{scenario and route}
\vspace{-0.3cm}
\end{figure}

The experiment is conducted indoors, specifically within an office environment. The actual scenario and the diagram of the route for data collection are illustrated in Fig. \ref{scenario and route}. The training and testing data are gathered along the blue route in Fig. \ref{scenario and route}(b). MATLAB is used to process the collected data, with the CSI extracted from valid data packets, where the CSI size is $3 \times 2$. For the sake of simplification, this paper focuses solely on the scenario of single-stream transmission, where $N_\mathrm{s} = 1$. In total, 50,000 samples have been obtained, which are divided into training, validation, and testing sets at a ratio of 8:1:1, respectively.

\subsection{Performance Metrics}

To characterize the accuracy of the reconstructed beamforming vector, we employ cosine similarity as the performance metric, which is computed as follows:
\begin{equation}
    \rho=\mathbb{E}\left\{\frac1{N_\mathrm{vs}}\sum_{n=1}^{N_\mathrm{vs}}\frac{|\hat{\mathbf{V}}_n \mathbf{V}_n|}{\|\hat{\mathbf{V}}_n\|_2\|\mathbf{V}_n\|_2}\right\},
\end{equation}
where $N_\mathrm{vs}$ is the number of valid subcarriers.

In addition, throughput is an important metric for evaluating system performance. The error vector magnitude (EVM) measures the errors between the measured symbols and expected symbols. In the calculation of throughput, the EVM of the received signal needs to be considered. In our experiment, we calculate EVM by simulating signal transmission and reception, where the transmission power is set to 30 dBm, and EVM of the received signal is calculated for each method. The selection of the modulation and coding scheme is contingent upon the EVM, which is calculated as follows:
\begin{equation}
    \mathrm{EVM}=10 \log_{10}{ \frac{\mathbb{E}\{ \|\hat{\mathbf{x}}[k]-\mathbf{x}[k]\|^2\}}{\mathbb{E}\{\|\mathbf{x}[k]\|^2\}} },~\text{(dB)}
\end{equation}
where $\mathbf{x}[k]$ and $\hat{\mathbf{x}}[k]$ are the original transmitted signal and the estimated received signal, respectively.

Gross throughput is the data rate achieved without consideration of CSI feedback. Based on EVM results, gross throughput is expressed by:
\begin{equation} \label{eq:gross_throughput}
    r=\frac{N_{\mathrm{vs}}}{N_{\mathrm{fft}}+N_{\mathrm{cp}}} \times b \times \gamma(\mathrm{EVM}),
\end{equation}
where $N_{\mathrm{vs}}$, $N_{\mathrm{fft}}$, and $N_{\mathrm{cp}}$ are the number of valid subcarriers, FFT points, and length of prefix length, respectively. In our experiment, they are set to 28, 64, and 16 respectively. $b$ is the sampling rate, set to 40 MSps.
The bandwidth and center frequency are 40 MHz and 2.4 GHz, respectively. $\gamma(\mathrm{EVM})$ is the average number of bits carried by one subcarrier, determined by $\mathrm{EVM}$\cite{IEEE802.11ac}.

Net throughput stands for the data rate achieved by the device after excluding the overhead caused by CSI feedback and similar factors, which can be expressed as:
\begin{equation} \label{eq:net_throughput}
    \bar{r}=\frac{T }{T+T_{\mathrm{overhead}}} \times r,
\end{equation}
where $T$ is the time for data transmission, $T_{\mathrm{overhead}}$ is the total transmission time of NDPA, NDP, ACK and CBR. The time allocated for CSI overhead, referred to as CBR time, will employ the BPSK rate considering feedback reliability. In our experiment, the length of data packet is 300 bytes.

\subsection{Results Analysis}

\begin{table}[!htb]
    \centering
    \renewcommand{\arraystretch}{1.3}  
    \caption{Cosine Similarity and Throughput Comparison of Each Method.}
    \resizebox{0.4\textwidth}{!}{
    \begin{tabular}{ccccc}
        \toprule
        Method  & T0  & T1  & LB-SciFi & EFNet \\
        \midrule
        Overhead & 672 & 896 & 102   & \textbf{100} \\
        Cosine similarity & 0.9991 & {\bf 0.9999} & 0.8786   & 0.9682 \\
        EVM (dB) & -17.85&	{\bf -18.06} & -12.29 & -13.54 \\
       Gross throughput (Mb/s) 
       & 28 & {\bf 28} & 14 & 21\\
       Net throughput (Mb/s) & 7.66 & 6.94 & 7.54 & \textbf{9.22}\\

    \bottomrule
\end{tabular}}

    \label{results_5300}
\end{table}
 
\begin{table*}[!htb]
    \centering
    \renewcommand{\arraystretch}{1.3}  
    \caption{Results of Each Method with OpenWiFi.}
    
    \begin{tabular}{ccccccccc}
        \toprule
        Method  & T0G1 & T0G2 & T0G4 & T1G1 & T1G2 & T1G4  & LB-SciFi & EFNet \\
        \midrule
        Overhead & 624 & 312 & 156 & 832 & 416 & 208 &136   & \textbf{120} \\
        Cosine similarity & 0.9995 & 0.9924 & 0.9755 & \textbf{0.9999} & 0.9929 & 0.9759 & 0.9885 & 0.9956 \\
        EVM (dB) & -17.98 & -16.77 & -13.28   & \textbf{-18.22} & -16.99 & -13.49 & -14.38 & -16.41 \\
        Gross throughput (Mb/s)
      & 26 & 26 & 19.5 & 26 & 26 & 19.5 & 19.5 & \textbf{26}\\
         Net throughput (Mb/s)&7.26&	8.50&	8.30&	6.62&	8.04&	8.07&	8.39	& \textbf{9.49}\\

        \bottomrule
    \end{tabular}

    \label{results_OpenWiFi}
\end{table*}

For ease of exposition, we use T0 and T1 to denote the IEEE 802.11 MIMO standard with Type 0 and Type 1 feedback, respectively. The CSI collected by the CSI Tool is sampled, implying that it has undergone subcarrier grouping. This process results in relatively weak correlation among the collected adjacent CSI; thus, no further grouping is performed. The performance comparison results of various methods are shown in Table \ref{results_5300}. Firstly, we compare the cosine similarity between the reconstructed $\hat{\mathbf{V}}$ and the original $\mathbf{V}$, as well as the feedback overhead for the standards, LB-SciFi \cite{sangdeh2020lb}, and EFNet. The results reveal that EFNet reduces the feedback overhead by 87.24\% on average compared to the standards, while maintaining a cosine similarity close to that of T0. This indicates that EFNet can effectively learn channel features with significantly fewer bits compared to the standard, achieving efficient resource utilization while ensuring high reconstruction accuracy.

Subsequently, we employ the reconstructed $\mathbf{V}$ for the simulation of signal transmission and reception to determine the EVM of the received signal. Then the throughput can be evaluated using \eqref{eq:gross_throughput} and \eqref{eq:net_throughput} for calculating throughputs, as depicted in Table \ref{results_5300}. The results indicate that EFNet achieves higher net throughput compared to T0, T1, and LB-SciFi.

Compared with LB-SciFi, the results show that EFNet can learn the optimal scheme for compression automatically, eliminating the necessity for manually designing the dimensions of the codewords and the quantization bits for $\mathbf{\Psi}$ and $\mathbf{\Phi}$.

To obtain results after subcarrier grouping, we also extracted the CSI by another means, namely, openWiFi \cite{jiao2020openwifi}. OpenWiFi can collect the CSI of consecutive subcarriers with strong correlation, and a total of 6,500 samples were collected for training the neural network.\footnote{It was found that OpenWiFi could not detect the signal from the specified AP in the same scenario as before. This may be due to OpenWiFi operating only in OFDM mode and supporting 802.11a/g; if the AP transmits a beacon in 802.11b modulation or the beacon is too recent, it cannot be detected by OpenWiFi.} The carrier frequency is set to 2.4 GHz, the channel bandwidth to 20 MHz, the number of valid subcarriers to 52, and the numbers of transmitting and receiving antennas are set to 2 and 1, respectively.
The experimental results are summarized in Table \ref{results_OpenWiFi}. The standard with Type 0/1 feedback and $1/2/4$ subcarriers in a group will henceforth be denoted as T0/1-G1/2/4. Results demonstrate that EFNet achieves reconstruction accuracy close to that of T0G1, with an 80.77\% reduction in feedback overhead and a 30.7\% increase in net throughput. Compared to LB-SciFi, EFNet achieves higher reconstruction accuracy, an 11.76\% reduction in feedback overhead, and a 13.11\% increase in net throughput.

\vspace{1mm}
\section{Conclusion}
\label{section:conclusion}
This paper introduces a DL-based CSI feedback framework named EFNet. Unlike traditional methods that require compressing the beamforming matrices into angles using Givens rotations, EFNet directly utilizes its encoder to compress the $\mathbf{V}$ matrices into codewords at the STA. These codewords are then quantized and reconstructed at the AP using the decoder. Compared to LB-SciFi, which necessitates preprocessing and manual design of quantization schemes, EFNet fully exploits the advantages of AI to achieve high-precision CSI reconstruction with significantly reduced feedback overhead. Experiments were conducted to validate the performance of our proposed approach. The results demonstrate that EFNet achieves high reconstruction accuracy with very low feedback overhead compared to both the standard protocol and LB-SciFi, while also delivering higher net throughput.

\bibliographystyle{IEEEtran}
\bibliography{reference}

\begin{thebibliography}{10}
\providecommand{\url}[1]{#1}
\csname url@samestyle\endcsname
\providecommand{\newblock}{\relax}
\providecommand{\bibinfo}[2]{#2}
\providecommand{\BIBentrySTDinterwordspacing}{\spaceskip=0pt\relax}
\providecommand{\BIBentryALTinterwordstretchfactor}{4}
\providecommand{\BIBentryALTinterwordspacing}{\spaceskip=\fontdimen2\font plus
\BIBentryALTinterwordstretchfactor\fontdimen3\font minus \fontdimen4\font\relax}
\providecommand{\BIBforeignlanguage}[2]{{%
\expandafter\ifx\csname l@#1\endcsname\relax
\typeout{** WARNING: IEEEtran.bst: No hyphenation pattern has been}%
\typeout{** loaded for the language `#1'. Using the pattern for}%
\typeout{** the default language instead.}%
\else
\language=\csname l@#1\endcsname
\fi
#2}}
\providecommand{\BIBdecl}{\relax}
\BIBdecl

\bibitem{IEEE802.11n}
{IEEE 802.11n}, ``{Part 11: Wireless LAN Medium Access Control (MAC) and Physical Layer (PHY) specifications. Amendment 5: Enhancement for Higher Throughput},'' 2009.

\bibitem{IEEE802.11ac}
{IEEE 802.11ac}, ``{Part 11: Wireless LAN Medium Access Control (MAC) and Physical Layer (PHY) Specifications. Amendment 4: Enhancements for Very High Throughput for Operation in Bands below 6GHz},'' 2013.

\bibitem{wen2018deep}
C.-K. Wen, W.-T. Shih, and S.~Jin, ``{Deep Learning for Massive MIMO CSI Feedback},'' \emph{IEEE Wireless Commun. Lett.}, vol.~7, no.~5, pp. 748--751, 2018.

\bibitem{yang2019deep}
Q.~Yang, M.~B. Mashhadi, and D.~G{\"u}nd{\"u}z, ``{Deep Convolutional Compression for Massive MIMO CSI Feedback},'' in \emph{Proc. 29th Int. Workshop Mach. Learn. Signal Process. (MLSP)}, 2019, pp. 1--6.

\bibitem{guo2022overview}
J.~Guo, C.-K. Wen, S.~Jin, and G.~Y. Li, ``{Overview of Deep Learning-Based CSI Feedback in Massive MIMO Systems},'' \emph{IEEE Trans. Commun.}, vol.~70, no.~12, pp. 8017--8045, 2022.

\bibitem{tang2022dilated}
S.~Tang, J.~Xia, L.~Fan, X.~Lei, W.~Xu, and A.~Nallanathan, ``{Dilated Convolution based CSI Feedback Compression for Massive MIMO Systems},'' \emph{IEEE Trans. Veh. Technol.}, vol.~71, no.~10, pp. 11\,216--11\,221, 2022.

\bibitem{bi2022novel}
X.~Bi, S.~Li, C.~Yu, and Y.~Zhang, ``{A Novel Approach Using Convolutional Transformer for Massive MIMO CSI Feedback},'' \emph{IEEE Wireless Commun. Lett.}, vol.~11, no.~5, pp. 1017--1021, 2022.

\bibitem{giordano2023will}
\BIBentryALTinterwordspacing
L.~G. Giordano, G.~Geraci, M.~Carrascosa, and B.~Bellalta, ``{What Will Wi-Fi 8 Be? A Primer on IEEE 802.11bn Ultra High Reliability},'' \emph{arXiv:2303.10442}, 2023. [Online]. Available: \url{https://arxiv.org/abs/2303.10442}
\BIBentrySTDinterwordspacing

\bibitem{wang2024}
\BIBentryALTinterwordspacing
X.~Wang, M.~Gan, and L.~Xin, ``{IEEE P802.11 - Artificial Intelligence Machine Learning (AIML) Topic Interest Group (TIG) - Meeting Update: Status of IEEE 802.11 Artificial Intelligence Machine Learning (AIML) TIG},'' 2024. [Online]. Available: \url{https://www.ieee802.org/11/Reports/aiml_update.htm}
\BIBentrySTDinterwordspacing

\bibitem{sangdeh2020lb}
P.~K. Sangdeh, H.~Pirayesh, A.~Mobiny, and H.~Zeng, ``{LB-SciFi: Online Learning-Based Channel Feedback for MU-MIMO in Wireless LANs},'' in \emph{Proc. IEEE 28th Int. Conf. Netw. Protocols (ICNP)}, 2020, pp. 1--11.

\bibitem{deshmukh2022intelligent}
M.~Deshmukh, M.~Kamel, Z.~Lin, R.~Yang, H.~Lou, and I.~G{\"u}ven{\c{c}}, ``{Intelligent Feedback Overhead Reduction (iFOR) in Wi-Fi 7 and Beyond},'' in \emph{Proc. IEEE 95th VTC-Spring}, 2022, pp. 1--5.

\bibitem{sangdeh2021deepmux}
P.~K. Sangdeh and H.~Zeng, ``{DeepMux: Deep-Learning-Based Channel Sounding and Resource Allocation for IEEE 802.11 ax},'' \emph{IEEE J. Sel. Area. Commun.}, vol.~39, no.~8, pp. 2333--2346, 2021.

\bibitem{10272483}
N.~Bahadori, Y.~Matsubara, M.~Levorato, and F.~Restuccia, ``[splitbeam: Effective and efficient beamforming in wi-fi networks through split computing],'' in \emph{Proc. IEEE 43rd ICDCS}, 2023, pp. 864--874.

\bibitem{IEEE802.11ax}
{IEEE 802.11ax}, ``{Part 11: Wireless LAN Medium Access Control (MAC) and Physical Layer (PHY) Specifications. Amendment 1: Enhancements for High Efficiency WLAN},'' 2019.

\bibitem{kim1999normalization}
D.~Kim, ``{Normalization Methods for Input and Output Vectors in Backpropagation Neural Networks},'' \emph{Int. J. Comput. Math.}, vol.~71, no.~2, pp. 161--171, 1999.

\bibitem{diederik2014adam}
D.~P. Kingma and J.~L. Ba, ``{Adam: A Method for Stochastic Optimization},'' in \emph{{Proc. Int. Conf. Learn. Represent. (ICLR)}}, 2015.

\bibitem{halperin2011tool}
D.~Halperin, W.~Hu, A.~Sheth, and D.~Wetherall, ``{Tool Release: Gathering 802.11n Traces with Channel State Information},'' \emph{ACM SIGCOMM Comput. Commun. Rev.}, vol.~41, no.~1, pp. 53--53, 2011.

\bibitem{jiao2020openwifi}
X.~Jiao, W.~Liu, M.~Mehari, M.~Aslam, and I.~Moerman, ``{Openwifi: A Free and Open-Source IEEE802. 11 SDR Implementation on SoC},'' in \emph{Proc. IEEE 91st VTC-Spring}, 2020, pp. 1--2.

\end{thebibliography}

\end{document}